\documentstyle[12pt,epsfig]{article}
\begin{document}
\setlength{\parskip}{0.45cm}
\setlength{\baselineskip}{0.75cm}
\begin{titlepage}
\begin{flushright}
DO-TH 96/23 \\ RAL-TR-96-097  \\ December 1996
\end{flushright}
\vspace{0.6cm}
\begin{center}
\Large
\hbox to\textwidth{\hss
{\bf Next-to-Leading Order Evolution} \hss}

\vspace{0.1cm}
\hbox to\textwidth{\hss
{\bf of Polarized and Unpolarized} \hss}

\vspace{0.1cm}
\hbox to\textwidth{\hss
{\bf Fragmentation Functions} \hss}

\vspace{1.2cm}
\large
M.\ Stratmann\\
\vspace{0.5cm}
\normalsize
Universit\"{a}t Dortmund, Institut f\"{u}r Physik, \\
\vspace{0.1cm}
D-44221 Dortmund, Germany \\
\vspace{1.2cm}
\large
W. Vogelsang \\
\vspace{0.5cm}
\normalsize
Rutherford Appleton Laboratory \\
\vspace{0.1cm}
Chilton, Didcot, Oxon OX11 0QX, England \\
\vspace{1.6cm}
{\bf Abstract} 
\vspace{-0.3cm}
\end{center}
We determine the two-loop 'time-like' Altarelli-Parisi splitting functions, 
appearing in the next-to-leading order $Q^2$-evolution equations for 
fragmentation functions, via analytic continuation of the corresponding 
'space-like' splitting functions for the evolution of 
parton distributions. We do this for 
the case of unpolarized fragmentation functions and - for the first time -
also for the functions 
describing the fragmentation of a longitudinally polarized parton into a 
longitudinally polarized spin-1/2 hadron such as a $\Lambda$ baryon. 
Our calculation is based on the method proposed and 
employed by Curci, Furmanski and Petronzio 
in the unpolarized case in which we confirm their results.
\end{titlepage}
%
%
%
\section{Introduction}
The spin structure of longitudinally polarized nucleons has been 
investigated in a number of experiments \cite{ref1} in recent years
by scattering polarized highly virtual 
space-like ($q^2\equiv -Q^2<0$) photons
off polarized targets, which provides access to the spin-dependent 
'space-like' parton distributions of the nucleon. In contrast to this,
nothing is known as yet about the corresponding polarized {\em 'time-like'} 
parton densities, i.e., the functions describing the fragmentation of a 
longitudinally polarized quark or gluon into a longitudinally polarized 
(spin-1/2) hadron. In analogy with the space-like case these are defined by 
\begin{equation}  \label{eq1}
\Delta D_f^h (z,Q^2) = D_{f(+)}^{h(+)}(z,Q^2) - D_{f(+)}^{h(-)}(z,Q^2) \; ,
\end{equation}
where $D_{f(+)}^{h(+)}(z,Q^2)$ ($D_{f(+)}^{h(-)}(z,Q^2)$) is the 
probability for finding a hadron $h$ with positive (negative) helicity in a 
parton $f$ with positive helicity at a mass scale $Q$, carrying a 
fraction $z$ of the parent parton's momentum. We note that taking the sum
instead of the difference in Eq.~(\ref{eq1}) one obtains the corresponding
unpolarized fragmentation function. 

Spin-dependent fragmentation functions 
appear equally interesting as the space-like 
distributions since they obviously
contain information on how the spin of a fragmenting parton is transmitted to 
that of the produced hadron. The most likely candidate for a measurement of 
polarized fragmentation functions is the $\Lambda$ baryon since its 
dominant decay $\Lambda \rightarrow p\pi^-$ is parity-violating and 
enables the determination 
of the $\Lambda$'s polarization \cite{perkins}. In \cite{bj} a strategy was
proposed for extracting the $\Delta D_f^{\Lambda}$ ($f=q,\bar{q}$)
in single-particle inclusive $e^+e^-$ annihilation (SIA) $e^+ e^- \rightarrow 
\Lambda X$. If the energy is far below the $Z$ resonance, one longitudinally 
polarized beam is required in order to fix the polarization of the outgoing 
(anti)quark that fragments into the $\Lambda$ and to obtain a non-vanishing
twist-two spin asymmetry. At higher energies, no beam polarization is needed 
since the parity-violating coupling $q\bar{q}Z$ automatically generates a
net polarization of the quarks. Apart from SIA, which plays the 
same fundamental role for the determination of fragmentation functions 
as deep-inelastic scattering (DIS) does for that of space-like parton
distributions, the possibility of extracting the $\Delta D_f^{\Lambda}$ 
in semi-inclusive DIS (SIDIS) in the current fragmentation region,
$ep \rightarrow e \Lambda X$, has also been studied theoretically recently 
\cite{jaf,luma}. Here, either a longitudinally polarized lepton beam or a 
polarized nucleon target would be required. On the experimental side, ALEPH 
has reported a first measurement of the $\Lambda$ polarization in $Z$ decays 
recently \cite{aleph}, and a measurement of the $\Delta D_f^{\Lambda}$ in 
SIDIS appears possible for the HERMES experiment \cite{court} and is planned 
by the COMPASS collaboration \cite{compass}.
              
The polarized cross section for, say\footnote{Since the $\Lambda$ appears
the most realistic candidate for measurements of spin-dependent fragmentation
functions we write down all formulae below for this specific case. 
They apply, of
course, equally well to any other final state (spin-1/2) hadron whose
longitudinal polarization can be determined experimentally.}, the process 
$e^+ \vec{e}^-\rightarrow \vec{\Lambda}X$ (the arrows denoting longitudinal 
polarization) can be written in the factorized form 
as a sum over convolutions 
of polarized hard subprocess 
cross sections with the {\em process-independent} 
fragmentation functions $\Delta D_f^{\Lambda}$ of (\ref{eq1}),
\begin{eqnarray} \label{wqdef}
\!\!\frac{d\Delta \sigma^{e^+ e^-\rightarrow \Lambda X}(s,y,z)}{dydz}      
&\equiv & \frac{d\sigma (\lambda_e =+,\lambda_{\Lambda}=+,s,y,z)}{dydz} -
\frac{d\sigma (\lambda_e =+,\lambda_{\Lambda}=-,s,y,z)}{dydz} \\ 
\label{wqdef1}
\!\!
&=& \sum_{f=q,\bar{q},g} \int_z^1 \frac{d\xi}{\xi} \frac{d\Delta 
\hat{\sigma}^{e^+ e^-\rightarrow f x} (s,y,\xi)}{dy d\xi} 
\Delta D_f^{\Lambda} (\frac{z}{\xi},Q^2) \; ,
\end{eqnarray}
where $\lambda_e$, $\lambda_{\Lambda}$ denote the helicities of the
polarized electron and the $\Lambda$, and $s\equiv 2 p_{e^+}\cdot p_{e^-}$,
$z\equiv 2 p_{\Lambda}\cdot q/Q^2$ with the momentum $q$ of the time-like
($q^2 \equiv Q^2 >0$) intermediate $\gamma$ or $Z$. The variable $y$ is 
defined by $y\equiv p_{\Lambda}\cdot p_{e^-}/p_{\Lambda}\cdot q$ and is 
related to 
the cms scattering angle $\theta$ of the produced $\Lambda$ with respect to 
the incoming electron via 
$y=(1+\cos \theta)/2$. The polarized subprocess cross sections 
$d\Delta \hat{\sigma}^{e^+ e^-\rightarrow f x} (s,y,\xi)/dy d\xi$ are 
defined in complete analogy with (\ref{wqdef}), and $\xi$ is the partonic 
counterpart of $z$, $\xi\equiv 2 p_{f} \cdot q/Q^2$. Again the corresponding 
expression for the unpolarized cross section is obtained by taking the sum 
instead of the difference in Eq.~(\ref{wqdef}) and omitting all $\Delta$'s. 
Unlike the fragmentation functions, the hard subprocess cross sections are 
calculable in perturbative QCD. QCD can, however, 
predict the $Q^2$ dependence 
of the fragmentation functions via the Altarelli-Parisi equations \cite{ap}, 
once a suitable non-perturbative hadronic input for the evolution has been 
found. In the leading order (LO), there is only one subprocess, namely 
$e^+ e^- \rightarrow q\bar{q}$ via $\gamma$ or $Z$ exchange, and the
fragmentation functions evolve according to the LO polarized (time-like)
Altarelli-Parisi equations. 

It is the main purpose of this paper to set up the complete next-to-leading 
order (NLO) framework of QCD for single-inclusive annihilation into a 
polarized hadron. This task first of all involves calculating the ${\cal O} 
(\alpha_s)$ corrections to the LO hard subprocess cross section, 
including the calculation of subprocesses that first appear at NLO. This was 
recently achieved in \cite{rav,arg}. However, knowledge of the underlying hard
subprocesses to NLO accuracy cannot be the full story as becomes immediately 
obvious from the well-known fact that the corresponding corrections are 
factorization scheme dependent, i.e., depend on the convention adopted when 
subtracting the collinear singularities appearing in the calculation. For a 
fully consistent NLO calculation one also needs to perform the evolution of 
the polarized fragmentation functions in NLO, which requires knowledge of the 
polarized NLO evolution kernels in the 
(time-like) Altarelli-Parisi equations. 
Only when both types of NLO corrections, those to the subprocess cross
sections {\em and} to the evolution kernels, are known does the NLO
framework become complete and consistent, the factorization scheme
dependencies cancelling out to the order considered whenever a physical 
cross section is calculated. This situation is of course completely
the same as in the more familiar space-like case of, e.g., DIS structure
functions.  

As will be discussed below, it is possible to derive the polarized NLO 
time-like evolution kernels by analytic continuation of their space-like 
counterparts which have been calculated recently \cite{vn,wv1,wv}. The
procedure for doing this has first been worked out for the unpolarized 
non-singlet 
case in \cite{cfp} and has also been used for the unpolarized singlet sector 
in \cite{fp}. We will pursue this method. The results we obtain refer to the 
$\overline{\rm{MS}}$ scheme and need to be combined with NLO corrections to 
the hard subprocess cross sections in the same scheme, as recently presented
for SIA and SIDIS in \cite{arg}. Since the procedure of analytic continuation 
can also be applied to the hard subprocess cross sections we will provide a 
check on the results of \cite{arg} for SIA.

In view of the present lack of any experimental information on the 
$\Delta D_f^{\Lambda}$ one could argue that it is somewhat premature to set 
up the full NLO framework for their evolution and the processes in which 
they appear. On the other hand, it seems likely that data will become 
available in the future. Furthermore, the transition from the space-like
to the time-like region in the polarized case appears interesting in itself:
In LO the space-like and time-like Altarelli-Parisi evolution kernels are 
related to each other via an analytic continuation rule (ACR) \cite{dly} and 
also via the so-called Gribov-Lipatov relation (GLR) \cite{gl}. In the 
unpolarized case these relations were shown to be broken beyond leading order 
in the $\overline{\rm{MS}}$ scheme \cite{cfp,fp,flor}, and a similar feature 
is thus expected for the spin-dependent case. The NLO effects also appear 
interesting from a more physical point of view. For instance, one would expect 
\cite{slambda} that to a first approximation polarized-$\Lambda$ production 
in SIA proceeds just via strange quark fragmentation $\vec{s}\rightarrow 
\vec{\Lambda}$, i.e., is essentially sensitive to $\Delta D_s^{\Lambda}$. 
NLO evolution on the other hand automatically generates non-vanishing 
non-strange fragmentation functions due to the existence of flavor 
non-diagonal quark-to-quark splitting functions. Also, the possibly important 
\cite{rav} role played by gluons is appreciated when going beyond the leading 
order.

The remainder of this paper is organized as follows. In section 2 we set the
general framework for our calculations and briefly discuss the LO results.
In sections 3 and 4 we present in some detail the determination of the NLO 
corrections for the unpolarized time-like situation by analytic continuation 
of their space-like counterparts. Even though neither the method of analytic
continuation nor the final result of the calculation are new, the full 
calculation itself has never been documented before, and we also provide 
new insight in the breakdown of the ACR beyond LO.
Furthermore, our findings 
in sections 3,4 are crucial for dealing with the polarized case, which is then
done in the subsequent section. In section 6 we study an interesting
supersymmetric relation obeyed by the NLO unpolarized and polarized time-like
splitting functions. Section 7 briefly summarizes our work. 
%
%
\section{General framework and LO results}
Let us first set the notation by collecting all 
ingredients for a NLO treatment 
of the cross section in Eq.~(\ref{wqdef}). We begin by dealing with the hard 
subprocess cross sections. In analogy with the familiar space-like 
$g_1 \equiv
g_1^{(S)}(x,Q^2)$ (where $x\equiv Q^2/2p\cdot q \leq 1$) 
we define a time-like 
structure function $g_1^{(T)} (z,Q^2)$ and write Eq.~(\ref{wqdef1}) 
as\footnote{For simplicity we restrict our considerations to pure photon 
exchange in the process $e^+ e^- \rightarrow q\bar{q}$. Exchange of $Z^0$ and 
$\gamma Z^0$ interference modify the angular dependence of the longitudinally 
polarized cross section and thus add new structure functions to its 
expression \cite{rav}.}
\begin{equation} \label{wqdef2}
\frac{d\Delta \sigma^{e^+ e^-\rightarrow \Lambda X}(s,y,z)}{dzdy}= 
\frac{6 \pi \alpha^2}{Q^2} (2y-1) g_1^{(T)} (z,Q^2) \; .
\end{equation}
To facilitate the further discussion, we adopt a combined treatment of the 
space-like and time-like situations and introduce the structure function
${\cal G}_1^{(U)}(\xi,Q^2)$, where the index $U$ 
stands for either 'space-like' 
($U=S$, ${\cal G}_1^{(S)} \equiv 2g_1^{(S)}$, $\xi=x$), 
or 'time-like' ($U=T$, 
${\cal G}_1^{(T)} \equiv g_1^{(T)}$, $\xi=z$), and the parton 
distributions $\Delta f^{(U)}(\xi,Q^2)$ ($f=q,\bar{q},g$), where 
$\Delta f^{(S)} \equiv \Delta f$ (with the usual polarized hadronic parton 
densities $\Delta f$) and 
$\Delta f^{(T)} \equiv \Delta D_f^\Lambda$. In terms 
of the $\Delta f^{(U)}$ we can write ${\cal G}_1^{(U)}$ to NLO as
\begin{equation} \label{g1}  
{\cal G}_1^{(U)} (\xi,Q^2) = \sum_{q} e_i^2\; 
\Bigg\{ \left[ \Delta q^{(U)} + \Delta \bar{q}^{(U)} \right] \otimes 
\Delta {\cal C}_q^{(U)} + \eta_U \Delta g^{(U)} 
\otimes \Delta {\cal C}_g^{(U)} \Bigg\} (\xi,Q^2) \; ,
\end{equation}
where the sum runs over the $n_f$ active quark flavors, $\eta_S=1/n_f$, 
$\eta_T=2$ and  $\otimes$ denotes the usual convolution.
The hard subprocess cross sections $\Delta {\cal C}_q^{(U)}$, 
$\Delta {\cal C}_g^{(U)}$ are taken to have the perturbative expansion
\begin{equation}
\Delta {\cal C}_i^{(U)}(\xi,\alpha_s) = \Delta C_i^{(U),(0)}(\xi) + 
\frac{\alpha_s}{2\pi} \Delta C_i^{(U),(1)} (\xi) \; ,
\end{equation}
where $\Delta C_q^{(U),(0)}(\xi)=\delta(1-\xi)$, 
$\Delta C_g^{(U),(0)}(\xi)=0$.

To determine the $Q^2$ evolution of the space-like and time-like parton 
densities $\Delta f^{(U)}$ in Eq.~(\ref{g1}) it is as usual convenient to 
decompose them into flavor singlet and non-singlet pieces by introducing the 
densities $\Delta q_{\pm}^{(U)}$ and the vector 
\begin{equation}
\Delta \vec{v}^{(U)}\equiv \left( \begin{array}{c} \Delta \Sigma^{(U)} \\
\Delta g^{(U)} \\ \end{array}\right) \; , 
\end{equation}
where 
\begin{equation} \label{nsdef}
\Delta q_{\pm}^{(U)} \equiv \Delta q^{(U)} 
\pm \Delta \bar{q}^{(U)} \; , \;\;\;
\Delta \Sigma^{(U)} \equiv \sum_q (\Delta q^{(U)}+\Delta \bar{q}^{(U)}) \; .
\end{equation}
One then has the following non-singlet 
evolution equations ($q,\tilde{q}$ being
two different flavors):
\begin{eqnarray} \label{ns1evol}
\frac{d}{d\ln Q^2} (\Delta q_+^{(U)} - \Delta \tilde{q}_+^{(U)}) (\xi,Q^2)&=& 
\left[\Delta P_{qq,+}^{(U)} \otimes (\Delta q_+^{(U)} - 
\Delta \tilde{q}_+^{(U)})\right] (\xi,Q^2) \; , \\
\label{ns2evol} 
\frac{d}{d\ln Q^2} \Delta q_-^{(U)} (\xi,Q^2) &=& 
\left[\Delta P_{qq,-}^{(U)} \otimes \Delta q_-^{(U)}\right] (\xi,Q^2) \;\;.  
\end{eqnarray}
The two evolution kernels $\Delta P_{qq,\pm}^{(U)}(\xi,\alpha_s(Q^2))$ 
start to become different 
beyond LO as a result of the presence of transitions between quarks 
and antiquarks. The singlet evolution equation reads
\begin{eqnarray} \label{sing}
\frac{d}{d\ln Q^2} \Delta \vec{v}^{(U)} (\xi,Q^2) = \left[ 
\Delta \hat{P}^{(U)}\otimes \Delta \vec{v}^{(U)}\right](\xi,Q^2) \; .
\end{eqnarray}
We write the singlet evolution matrices for the space-like and time-like  
cases as
\renewcommand{\arraystretch}{1.3}
\begin{eqnarray} \label{singmat}
\Delta \hat{P}^{(S)} \equiv \left( \begin{array}{cc} 
\Delta P_{qq}^{(S)} &  \Delta P_{qg}^{(S)} \\  
\Delta P_{gq}^{(S)} &  \Delta P_{gg}^{(S)} \\
\end{array}\right) \; , \;\;
\Delta \hat{P}^{(T)} \equiv \left( \begin{array}{cc} 
\Delta P_{qq}^{(T)} &  2n_f \Delta P_{gq}^{(T)} \\  
\frac{1}{2n_f} \Delta P_{qg}^{(T)} &  \Delta P_{gg}^{(T)} \\
\end{array}\right) \; . 
\end{eqnarray}  
The $qq$-entries in (\ref{singmat}) are expressed as
\begin{equation}  \label{qqs}
\Delta P_{qq}^{(U)}=\Delta P_{qq,+}^{(U)} + \Delta P_{qq,PS}^{(U)} \; .
\end{equation}
$\Delta P_{qq,PS}^{(U)}$ which vanishes in LO is called the 'pure singlet' 
splitting function since it only appears in the singlet case. 
To NLO, all splitting functions in (\ref{ns1evol})-(\ref{qqs}) have the 
perturbative expansion
\begin{equation} \label{expan}
\Delta P_{ij}^{(U)} (\xi,\alpha_s) = \left( \frac{\alpha_s}{2\pi} \right) 
\Delta P_{ij}^{(U),(0)} (\xi) + \left( \frac{\alpha_s}{2\pi} \right)^2 
\Delta P_{ij}^{(U),(1)} (\xi) \; .
\end{equation}

Just like their unpolarized counterparts, the polarized space-like and 
time-like splitting functions are equal in LO:
\begin{equation} \label{GLR}
\Delta P_{ij}^{(T),(0)} (\xi) = \Delta P_{ij}^{(S),(0)} (\xi) \; .
\end{equation}
Eqs.~(\ref{GLR}) are manifestations of the so-called Gribov-Lipatov relation 
(GLR) \cite{gl} which connects space-like and time-like structure functions
within their respective physical regions $(\xi<1)$ and is known to be 
broken beyond LO in the unpolarized case \cite{cfp,flor}. 
Recalling that for $x<1$ \cite{ap,ar}
\begin{eqnarray} \label{losplit}
\Delta P_{qq}^{(S),(0)}(x)&=&C_F \frac{1+x^2}{1-x} \; , \nonumber  \\
\Delta P_{qg}^{(S),(0)}(x)&=&2 T_f\left[ 2 x-1 \right]  \; , \nonumber \\
\Delta P_{gq}^{(S),(0)}(x)&=&C_F \left[ 2-x \right] \; , \nonumber \\
\Delta P_{gg}^{(S),(0)}(x)&=&2 C_A \Big[ \frac{1}{1-x}-2 x+1 \Big] \; ,
\end{eqnarray}
where 
\begin{equation}
C_F=\frac{4}{3},\; C_A=3,\; T_f=T_R n_f = \frac{1}{2} n_f, \; 
\beta_0 = \frac{11}{3} C_A -\frac{4}{3} T_f \;, 
\end{equation}
it becomes obvious that the space-like and time-like LO quantities
are also directly related by analytic continuation through $x=1$:
\begin{eqnarray} \label{ACR}
&&\Delta P_{qq,\pm}^{(T),(0)}(z) = -z \Delta P_{qq,\pm}^{(S),(0)} 
(\frac{1}{z}) \; , \nonumber \\
&&\Delta P_{qq}^{(T),(0)}(z) = -z \Delta P_{qq}^{(S),(0)} (\frac{1}{z}) \; ,
\; \; 
\Delta P_{gq}^{(T),(0)}(z) =  \frac{C_F}{2 T_f} z \Delta P_{qg}^{(S),(0)} 
(\frac{1}{z}) \; ,
\nonumber \\
&&\Delta P_{qg}^{(T),(0)}(z) =  \frac{2 T_f}{C_F} z \Delta P_{gq}^{(S),(0)} 
(\frac{1}{z}) \; , \; \; 
\Delta P_{gg}^{(T),(0)}(z) = -z \Delta P_{gg}^{(S),(0)} (\frac{1}{z}) \; ,
\end{eqnarray}
where $z<1$. For future convenience we have explicitly written out the  
singlet as well as the non-singlet sector even though all LO quark-to-quark 
splitting functions coincide, $\Delta P_{qq,+}^{(U),(0)}=\Delta 
P_{qq,-}^{(U),(0)} =\Delta P_{qq}^{(U),(0)}$. Eqs.~(\ref{ACR}) represent the 
analytic continuation or Drell-Levy-Yan relation (ACR) \cite{dly} to LO which
we cast into the generic form
\begin{equation} \label{ACRgen}
\Delta P_{ij}^{(T),(0)}(z) = z {\cal AC} \Bigg[ \Delta P_{ji}^{(S),(0)} 
(x=\frac{1}{z}) \Bigg] \; ,
\end{equation}
where the operation ${\cal AC}$ analytically continues any function to 
$x \rightarrow 1/z >1$ and correctly adjusts the color factor and the sign 
depending on the splitting function 
under consideration, cf.\ Eqs.~(\ref{ACR}). 
The LO relations (\ref{ACR}) are based on symmetries of tree diagrams under 
crossing, and one therefore has to expect that they are in general no longer 
valid when going to NLO, depending on the regularization and the 
factorization/renormalization prescriptions used in the NLO calculation. This 
is exactly what happens in dimensional regularization in the 
$\overline{\rm{MS}}$ scheme as was shown in \cite{cfp} for the unpolarized 
non-singlet case. Fortunately, as was also demonstrated in \cite{cfp}, the 
breaking of the ACR arising beyond LO is essentially due to kinematics and 
can therefore be rather straightforwardly detected within the method used 
in \cite{cfp,fp,ev} to calculate splitting functions. We will now first 
collect the findings of \cite{cfp} concerning the connection between the 
space-like and time-like flavor non-singlet configurations in the unpolarized 
case and analyze in detail their extension to the singlet sector made in
\cite{fp}. Afterwards we will apply the 
results to the polarized case. 
\section{NLO results for the unpolarized case}
Eqs.~(\ref{wqdef2})-(\ref{ACRgen}) above have been written down for the 
polarized case, but they all apply equally well to the unpolarized one when 
all $\Delta$'s are removed and, obviously, the unpolarized LO splitting
functions $P_{ij}^{(S),(0)}$ as calculated in \cite{ap},
\begin{eqnarray} \label{ulosplit}
P_{qq}^{(S),(0)}(x)&=&C_F \frac{1+x^2}{1-x} \; , \nonumber  \\
P_{qg}^{(S),(0)}(x)&=&2 T_f\left[ x^2+(1-x)^2 \right]  \; , \nonumber \\
P_{gq}^{(S),(0)}(x)&=&C_F \left[ \frac{1+(1-x)^2}{x} \right] \; , 
\nonumber \\
P_{gg}^{(S),(0)}(x)&=&2 C_A \Big[ \frac{x}{1-x}+\frac{1-x}{x}+
x (1-x) \Big] \; ,
\end{eqnarray}
(for $x<1$), are used in Eq.~(\ref{losplit}). Furthermore, in the unpolarized 
case with pure photon exchange one needs to introduce two independent 
structure functions ${\cal F}_1^{(U)}$, ${\cal F}_2^{(U)}$ (see, e.g., 
\cite{aem,aemp,fpzp}) with short-distance cross sections 
${\cal C}_{i,1}^{(U)}$, ${\cal C}_{i,2}^{(U)}$ ($i=q,g$), respectively.  

In \cite{cfp,fp,ev} the unpolarized NLO evolution kernels for the 
space-like situation were calculated using a method \cite{egmpr} that is
as close as possible to parton model intuition since it is based
explicitly on the factorization properties of mass singularities 
in the light-like axial gauge. The general strategy here consists of a 
rearrangement of the perturbative expansion which makes explicit the 
factorization into a part which does not contain any mass singularity and 
another one which contains all 
(and only) mass singularities. More explicitly, 
$M_{j,k}$ ($j=q,g$, $k=1,2$), the contribution of virtual (space-like) 
photon--quark or photon--gluon scattering to the structure functions 
${\cal F}_k^{(S)}$ on parton-level, is expanded into two-particle 
irreducible (2PI) kernels. 
In the light-cone gauge these 2PI kernels have been 
proven to be finite as long as the external legs are kept unintegrated, such 
that all collinear singularities originate from the integrations over the 
momenta flowing in the 
lines connecting the various kernels \cite{egmpr}. This 
allows for projecting out these singularities \cite{cfp}, and $M_{j,k}$ can 
thus be written in the factorized form              
\begin{equation}
M_{j,k} = \sum_{i=q,g} {\cal C}_{i,k}^{(S)} \otimes \Gamma_{ij}^{(S)} \;,
\end{equation}
where the ${\cal C}_{i,k}^{(S)}$ are finite (and obviously depend on the 
hard process considered), whereas the $\Gamma_{ij}^{(S)} \equiv 
\Gamma_{ij}^{(S)} (x,\alpha_s,1/\epsilon)$ 
contain just the mass singularities 
(which appear as poles in $\epsilon$ when using dimensional regularization, 
$d=4-2 \epsilon$) and are process-{\em in}dependent. The $\Gamma_{ij}^{(S)}$ 
are to be convoluted with bare ('unrenormalized') parton densities which 
must cancel their poles. As was shown in \cite{cfp}, the
$\overline{\rm{MS}}$ scheme Altarelli-Parisi \cite{ap} kernels one is looking 
for, appear order by order as the residues of the $1/\epsilon$ poles in 
$\Gamma_{ij}^{(S)}$, 
\begin{equation} \label{gamap}
\Gamma^{(S)}_{ij} (x,\alpha_s,\frac{1}{\epsilon} )=
\delta (1-x) \delta_{ij} - \frac{1}{\epsilon} \left[ \left(\frac{\alpha_s}
{2 \pi} \right) P_{ij}^{(S),(0)} (x) + \frac{1}{2} \left(  
\frac{\alpha_s}{2 \pi}\right)^2 P_{ij}^{(S),(1)} (x) + ... \right] + 
{\cal O} (\frac{1} {\epsilon^2} ) \; .
\end{equation}
The NLO contribution to the hard short-distance cross sections in the 
$\overline{\rm{MS}}$ scheme is obtained by calculating the full ('bare') 
subprocess cross sections $\hat{C}_{i,k}^{(S),(1)} (x,\frac{1}{\epsilon})$ 
($i=q,g$, $k=1,2$) and subtracting off the poles:
\begin{equation} \label{cpole}
C_{i,k}^{(S),(1)} (x) = \hat{C}_{i,k}^{(S),(1)} (x,\frac{1}{\epsilon}) + 
\frac{1}{\epsilon} \left( \frac{Q^2}{4\pi \mu^2} \right)^{-\epsilon} 
\frac{\Gamma (1-\epsilon)}{\Gamma (1-2 \epsilon)} P_{qi}^{(S),(0)}(x) \; ,  
\end{equation}
where $\mu$ is the arbitrary mass scale to be introduced in dimensional 
regularization. 

In the time-like region one can repeat the above procedure and introduce 
analogous quantities $\Gamma_{ji}^{(T)}(z,\alpha_s,1/\epsilon)$ that contain 
all final-state mass singularities arising in a fragmentation process. It 
turns out \cite{cfp} that the task of establishing the connection between 
$\Gamma_{ij}^{(S)}$ and $\Gamma_{ji}^{(T)}$ via analytic continuation can be 
reduced to understanding the differences between the 2PI kernels in the 
space-like and time-like situations. These essentially amount \cite{cfp} to 
relative extra phase space factors of 
$(k_2\cdot n/k_1 \cdot n)^{-2 \epsilon}$ 
in the time-like case, where $k_1$ and $k_2$ are the momenta of the particles 
entering or leaving a 2PI kernel, respectively. Here $n$ is the vector 
specifying the light-cone gauge and the longitudinal direction, i.e., 
$(k_2\cdot n/k_1 \cdot n)^{-2 \epsilon}
\equiv \zeta^{-2\epsilon}$ with $\zeta$ 
to be interpreted as the fraction of the 
momentum $k_1$ transferred to the particle with $k_2$. In the unpolarized case
a further difference arises from the spin-average factor for initial-state 
gluons which is $(d-2)^{-1}=1/2(1-\epsilon)$ in $d$ dimensions. As apparent 
from (\ref{ACR}), the off-diagonal splitting functions 
interchange their roles 
during the transition from the space-like to the time-like situation. 
In particular, the space-like $P_{qg}^{(S)}$ 
which includes the spin-averaging 
factor $(d-2)^{-1}$ gives rise to the time-like $P_{gq}^{(T)}$ which should 
just have the spin-average $1/2$, and vice versa for $P_{gq}^{(S)}$, 
$P_{qg}^{(T)}$. These effects have to be taken into account along with those 
coming from the 
$(k_2\cdot n/k_1 \cdot n)^{-2 \epsilon}$ terms mentioned above.
Consequently, all this gives on aggregate for $z<1$
\begin{eqnarray} \label{ACRNLO}
&&\Gamma_{qq,\pm}^{(T)}(z,\alpha_s,\frac{1}{\epsilon}) = -z^{1-2 \epsilon} 
\Gamma_{qq,\pm}^{(S)} (\frac{1}{z},\alpha_s,\frac{1}{\epsilon}) \; , 
\nonumber \\
&&\Gamma_{qq}^{(T)}(z,\alpha_s,\frac{1}{\epsilon}) = -z^{1-2 \epsilon} 
\Gamma_{qq}^{(S)} (\frac{1}{z},\alpha_s,\frac{1}{\epsilon}) \; , \; \; 
\Gamma_{gq}^{(T)}(z,\alpha_s,\frac{1}{\epsilon}) =  \frac{C_F}{2 T_f} 
z^{1-2 \epsilon} (1-\epsilon) \Gamma_{qg}^{(S)} 
(\frac{1}{z},\alpha_s,\frac{1}{\epsilon}) \; ,
\nonumber \\
&&\Gamma_{qg}^{(T)}(z,\alpha_s,\frac{1}{\epsilon}) =  \frac{2 T_f}{C_F} 
\frac{z^{1-2 \epsilon}}{1-\epsilon} \Gamma_{gq}^{(S)} 
(\frac{1}{z},\alpha_s,\frac{1}{\epsilon}) \; , \; \; 
\Gamma_{gg}^{(T)}(z,\alpha_s,\frac{1}{\epsilon}) = -z^{1-2 \epsilon}     
\Gamma_{gg}^{(S)} (\frac{1}{z},\alpha_s,\frac{1}{\epsilon}) \; .
\end{eqnarray}
We also include now the corresponding relations for the hard subprocess cross 
sections $\hat{C}_{i,k}^{(U),(1)}$ {\em before} subtraction of their pole 
terms (see Eq.~(\ref{cpole})):
\begin{eqnarray} 
\hat{C}_{q,k}^{(T),(1)} (z,\frac{1}{\epsilon}) &=& -z^{1-2 \epsilon}     
\hat{C}_{q,k}^{(S),(1)} (\frac{1}{z},\frac{1}{\epsilon}) \; , \nonumber \\
\hat{C}_{g,k}^{(T),(1)} (z,\frac{1}{\epsilon}) &=& \frac{C_F}{2 T_f} 
z^{1-2 \epsilon} (1-\epsilon) \hat{C}_{g,k}^{(S),(1)} 
(\frac{1}{z},\frac{1}{\epsilon}) \; . \label{ACRNLOc}     
\end{eqnarray}
It becomes obvious that higher pole terms in the expression for 
$\Gamma_{ij}^{(S)}$ in (\ref{gamap}) will generate additional contributions 
to the single pole of $\Gamma_{ji}^{(T)}$ when they are combined with the 
factors $z^{-2 \epsilon}$ or $(1-\epsilon)^{\pm 1}$ in (\ref{ACRNLO}), e.g.,
\begin{equation}
\frac{1}{\epsilon^2} z^{-2 \epsilon} = \frac{1}{\epsilon^2} 
-\frac{2}{\epsilon} \ln z + {\cal O} (1) \; . 
\end{equation}
In the same way the pole terms in $\hat{C}_{i,k}^{(S),(1)}(x,
\frac{1}{\epsilon})$ will give rise to extra finite contributions to the
$\hat{C}_{i,k}^{(T),(1)}$ that remain after the pole is subtracted.
Following \cite{cfp} we separate all such ACR-violating 
contributions by writing
\begin{eqnarray} \label{ACRNLO1}
\Gamma_{ij}^{(T)} (z,\alpha_s,\frac{1}{\epsilon}) &=&  z {\cal AC} \Bigg[ 
\Gamma_{ji}^{(S)} (x=\frac{1}{z},\alpha_s,\frac{1}{\epsilon}) \Bigg] +
\Gamma_{ij}^{\epsilon}(z,\alpha_s,\frac{1}{\epsilon}) \; , \nonumber \\
\hat{C}_{i,k}^{(T),(1)} (z,\frac{1}{\epsilon}) &=& z {\cal AC} \Bigg[ 
\hat{C}_{i,k}^{(S),(1)} (x=\frac{1}{z},\frac{1}{\epsilon}) \Bigg] +
C_{i,k}^{\epsilon,(1)} (z) \; , 
\end{eqnarray}
where, as before, $k=1,2$, $i,j=q,g$ (or '$ij=qq,\pm$' for the non-singlet 
case). We have extended the notation ${\cal AC}[...]$ for the analytic 
continuation (see Eq.~(\ref{ACRgen})) to the short-distance cross sections, 
its action here being obvious from Eq.~(\ref{ACRNLOc}).     
\begin{figure}[htb]
\vspace*{-0.5cm}
\epsfig{file=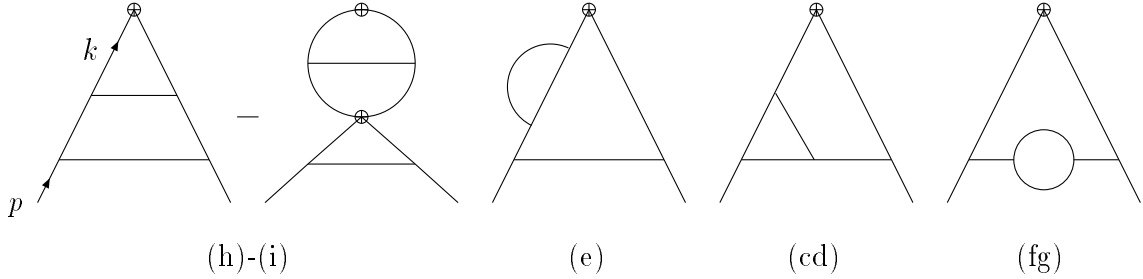,width=17cm}
\vspace*{-7cm}
\caption{Basic topologies of the diagrams that contribute to 
$\Gamma_{ij}^{\epsilon}$ (as defined in Eq.~(\ref{ACRNLO1})) in NLO.}
\end{figure}
One can now go through the NLO calculation \cite{ev} of the 
$\Gamma_{ij}^{(S),(1)}$ graph by graph to pick up the $1/\epsilon^2$ pole 
terms and thus to extract the contributions to the $\Gamma_{ij}^{\epsilon}
(z,\alpha_s,1/\epsilon)$ that break the ACR. For this purpose we present
in Fig.~1 the basic topologies for all NLO diagrams involved here,
where the notation is as introduced in \cite{cfp,ev}\footnote{We note that
the remaining topologies ((b),(jk)) introduced in \cite{cfp,ev} do not
possess higher pole terms and thus do not contribute to the 
$\Gamma_{ij}^{\epsilon}$.}. 
For topologies (cd),(e),(fg) the higher pole terms 
are necessarily proportional to the pole terms in 
the renormalization constants
as listed in \cite{cfp,ev}. The corrections to the 
ACR coming from these graphs
are therefore easily detected upon 
insertion into Eq.~(\ref{ACRNLO}). The higher 
pole terms of the genuine ladder graph (h), which is quadratic in the LO 2PI 
kernels, are given by the convolution of the LO splitting function for the 
upper rung with that corresponding to the lower one, and again the result is 
straightforwardly obtained. The only subtlety arises for the subtraction 
graphs (i) whose contributions to $\Gamma_{ij}^{(S),(1)}$ are proportional to 
\cite{wv1}
\begin{equation} \label{icon}
{\rm graph (i)} \sim \frac{1}{\epsilon^2} \Bigg( (1-z)^{-\epsilon} 
P_{ik,4-2 \epsilon}^{(S),(0)} \Bigg) \otimes P_{kj}^{(S),(0)} \; ,
\end{equation}
where $P_{ik,4-2\epsilon}^{(S),(0)}$ 
denotes the $(d=4-2 \epsilon)$-dimensional
LO splitting function standing for the upper part of the diagram, the factor 
$(1-z)^{-\epsilon}$ arising from phase space. The lower part of graph (i) 
is represented by $P_{kj}^{(S),(0)}$ which is the usual {\em four}-dimensional
LO splitting function. Application of the rules (\ref{ACRNLO}) to 
topology (i) is then to be understood as to include the kinematical 
$z^{-2 \epsilon}$ corrections only in the kernel representing the {\em upper} 
part of the diagram \cite{cfp}, which gives
\begin{equation}
{\rm graph (i)} \sim z {\cal AC} \Bigg[ \frac{1}{\epsilon} \Bigg( -2\ln z 
P_{ik}^{(S),(0)} \Bigg) \otimes P_{kj}^{(S),(0)} \Bigg] 
\end{equation}
as the contributions to the $\Gamma_{ij}^{\epsilon}$. If required, the 
spin-averaging factors $(1-\epsilon)^{\pm 1}$ also have to be taken into 
account. Contrary to all other topologies, adjusting the spin-averaging in 
graphs (i) generates corrections also to the ACR for 
the {\em diagonal} NLO singlet splitting functions $P_{qq,PS}^{(U),(1)}$ and 
$P_{gg}^{(U),(1)}$. As an example, Fig.~2 shows the graph of topology (i)
\begin{figure}[htb]
\vspace*{-1cm}
\epsfig{file=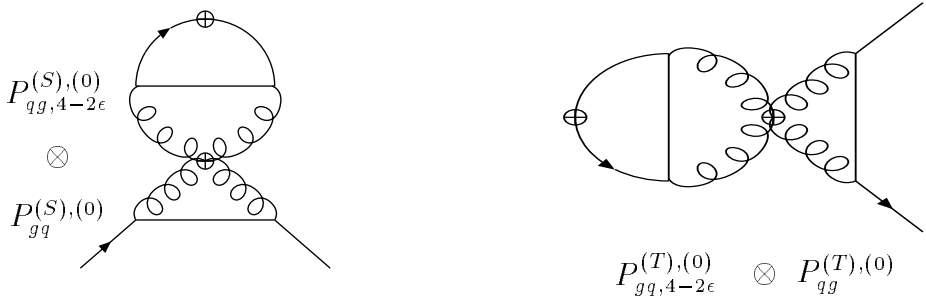,width=22cm,height=15cm}
\vspace*{-10cm}
\caption{Graphs of topology (i) contributing to $P_{qq,PS}^{(U),(1)}$ in the 
space-like (left) and time-like (right) situations.}
\end{figure}
for the case of the 'pure singlet' function $P_{qq,PS}^{(U),(1)}$ for both the
space-like and the time-like situations. According to Eq.~(\ref{icon}), the
contribution of this graph in the space-like case involves the 
convolution of the $d$-dimensional splitting function 
$P_{qg,4-2 \epsilon}^{(S),(0)}$ -- which contains the spin-averaging factor 
$(1-\epsilon)^{-1}$ -- with the {\em four}-dimensional function 
$P_{gq}^{(S),(0)}$. In the time-like case this obviously turns into the 
convolution of the $d$-dimensional $P_{gq,4-2\epsilon}^{(T),(0)}$ with the 
{\em four}-dimensional $P_{qg}^{(T),(0)}$, and {\em no} factor 
$(1-\epsilon)^{-1}$ is involved. The reversed situation appears in the $C_F 
T_f$ part of $P_{gg}^{(U),(1)}$. It is straightforward to account for these 
effects. 

We have now collected all ingredients for calculating the corrections to
the ACR arising from the $z^{-2\epsilon}$ terms (and, if applicable, the 
spin-averaging factors) in Eqs.~(\ref{ACRNLO}), i.e., for the functions 
$\Gamma_{ij}^{\epsilon}$ and $C_{i,k}^{\epsilon,(1)}$ in (\ref{ACRNLO1}). 
Following Eqs.~(\ref{gamap}),(\ref{cpole}) we keep just the residues of the 
$1/\epsilon$ poles in $\Gamma_{ij}^{\epsilon}$ and subtract the pole terms 
from the 'bare' subprocess cross sections. We then rewrite Eq.~(\ref{ACRNLO1})
as 
\begin{eqnarray} \label{ACRNLO2}
P_{ij}^{(T),(1)} (z) &=&  z {\cal AC} \Bigg[ P_{ji}^{(S),(1)} (x=\frac{1}{z}) 
\Bigg] + P_{ij}^{\epsilon,(1)}(z) \; , \\
C_{i,k}^{(T),(1)} (z) &=& z {\cal AC} 
\Bigg[ C_{i,k}^{(S),(1)} (x=\frac{1}{z}) 
\Bigg] + C_{i,k}^{\epsilon,(1)} (z) \; , \label{ACRNLO2c}
\end{eqnarray}
where, using the unpolarized LO splitting functions \cite{ap} of 
Eq.~(\ref{ulosplit}), we have
\begin{eqnarray} \label{unpdiff}
P_{qq,\pm}^{\epsilon,(1)}(z) &=& \beta_0 P_{qq}^{(S),(0)}(z)\ln z \;\; , 
\nonumber \\                     
P_{qq,PS}^{\epsilon,(1)}(z) &=& -C_F T_f \frac{4}{9z}\left[(1 - z) (38 + 47 z 
+ 38 z^2 ) +3 (1+z)(4+11z+4z^2)\ln z\right] \; , \nonumber \\
P_{gq}^{\epsilon,(1)}(z) &=& 2 (C_F-C_A) P_{gq}^{(S),(0)} (z) 
\Bigg[ -4 S_1(z)+2 \ln(1-z)\ln z + \ln(1-z) \Bigg] \nonumber \\
&+& \frac{C_F^2}{2z} \Bigg[ -4z (3-z) \ln z -20 +28z-5z^2 \Bigg] \nonumber \\
&+& \frac{C_A C_F}{9z} \Bigg[ 6 \left(4z^3+15z^2+18z+19\right)\ln z + 
(1-z) (235+55z+64 z^2) \Bigg] \;\; , \nonumber \\
P_{qg}^{\epsilon,(1)}(z) &=& \frac{4}{3} T_f P_{qg}^{(S),(0)} (z) (1-2\ln z)
\nonumber \\
&+& 2 (C_F-C_A) P_{qg}^{(S),(0)} (z) \Bigg[ 4 S_1(z)-2 \ln(1-z)\ln z+
\ln(1-z) \Bigg] \nonumber \\
&+& C_F T_f \left[ - 4 (2-z)(1-2z)\ln z -5+28z-20 z^2 \right] \nonumber \\
&+& \frac{2C_A T_f}{9z} \Bigg[ 6(-4-z-46z^2+9z^3) \ln z -64 -24z-114z^2+
169z^3 \Bigg] \;\; , \nonumber \\
P_{gg}^{\epsilon,(1)}(z) &=& - P_{qq,PS}^{\epsilon,(1)}(z) 
+ \beta_0 P_{gg}^{(S),(0)}(z) \ln z \;\; , \nonumber \\
C_{q,k}^{\epsilon,(1)} (z) &=& 2 \ln z P_{qq}^{(S),(0)}(z) \; , \nonumber \\
C_{g,k}^{\epsilon,(1)} (z) &=& (2 \ln z+1) P_{gq}^{(S),(0)}(z) \; .
\end{eqnarray}
For the off-diagonal $P_{ij}^{\epsilon,(1)}$ in (\ref{unpdiff}) we have 
introduced the function \cite{fp}
\begin{equation}
S_1 (z) \equiv \int_0^{1-z}  dy \frac{\ln (1-y)}{y} = -\mbox{Li}_2 (1-z)
\end{equation}
with the Dilogarithm function $\mbox{Li}_2$. 

The final step is to determine the analytic continuation of the space-like NLO 
splitting functions $P_{ji}^{(S),(1)}$ as published in \cite{cfp,fp,ev} and of 
the short-distance cross sections $C_{i,k}^{(S),(1)}$ (see, e.g., \cite{fpzp}) 
by using the operation ${\cal AC} [...]$ defined above. 
This is a straightforward task apart from two subtleties 
\cite{cfp}. Firstly, one has to recall that -- as a result of the finiteness 
of the 2PI kernels in the light-cone gauge -- the $1/\epsilon$ poles in the 
expression for $\Gamma^{(S)}_{ji}$ originate from the final integration over 
the momentum $k$ of the off-shell particle emerging from the uppermost kernel 
(see Fig.~1 for notation). This momentum obviously satisfies $k^2<0$ in the 
space-like case, but becomes {\em time-like} ($k^2>0$) when dealing with the 
$\Gamma^{(T)}_{ij}$. In other words, the transition from the space-like 
to the time-like situation does not only involve analytically continuing
to $x>1$ but also to $k^2>0$, crossing the threshold at $k^2=0$. Since 
the $1/\epsilon$ poles arise from terms like $(-k^2)^{-1-\epsilon}$ (for the 
space-like case) which are integrated over $-k^2$ down to $-k^2=0$, 
factors of $(-1)^{-\epsilon}$ will appear in the virtual integrals when going 
to $k^2>0$. Using ${\rm Re}[(-1)^{-\epsilon}]\approx 1-\epsilon^2 \pi^2/2$ one
realizes that this will result in extra $\pi^2$-contributions when multiplied 
by double pole terms present in intermediate stages of the calculations 
\cite{cfp}. The latter arise from the emission of soft gluons in the $qqg$
vertex. A similar feature is present for the $qq\gamma$ vertex \cite{aem} and
thus appears in the quark short-distance cross section $C_{q,k}^{(U),(1)}$. 
Here it is the transition $q^2<0 \rightarrow q^2>0$ 
that is responsible for the 
effect \cite{cfp} which only affects the endpoint contributions at $z=1$ to
be discussed below. 

The other subtlety concerns the analytic continuation to $x>1$ of terms $\sim
\ln^i (1-x)$ $(i=1,2)$ appearing in the $P_{ji}^{(S),(1)}(x)$ and in the
$C_{i,k}^{(S),(1)}$. To find the correct answer for this, one has to go 
through the relevant real (phase-space), 
virtual, and convolution integrals in 
the limit $x>1$. It turns out (see also \cite{cfp}) that all integrals can be 
smoothly continued through $x=1$ via $\ln^i(1-x) \rightarrow \ln^i(x-1)$, the 
only relevant exception being one of the scalar three-point functions with a 
light-cone gauge propagator which, for $x<1$, was given in Eq.~(A.15) of 
ref.~\cite{ev}. This particular three-point function only contributes
to the NLO splitting functions, but not to the hard subprocess cross sections
$C_{q,k}^{(S),(1)}$. Upon recalculation of the function for $x>1$ one finds 
that the correct continuation yields $\ln^2(1-x) \rightarrow \ln^2 (x-1) + 
\pi^2$ in this case. When combining this result with that for the
crossing of the $k^2=0$ threshold discussed above, one arrives at the 
interesting finding that the correct analytic continuation to $x>1$ of 
{\em all} terms $\sim \ln^i (1-x)$ $(i=1,2)$ in the NLO {\em splitting 
functions} $P_{ji}^{(S),(1)}(x)$ 
is effectively obtained by simply substituting 
\begin{eqnarray} \label{ln1}
\ln \left( 1-x \right) &\longrightarrow& \ln \left( x-1 \right) \; , \\
\ln^2 \left( 1-x \right) &\longrightarrow& \ln^2 \left( x-1 \right) - \pi^2 
\; .
\end{eqnarray}
For the non-singlet case, in which no $\ln^2 (1-x)$ terms appear in the
space-like NLO splitting function, this result is in agreement with the 
conclusion drawn in ref.~\cite{cfp} that the extra $\pi^2$ terms stemming
from the threshold at $k^2=0$ and from the three-point function cancel 
each other. In case of the NLO short-distance cross sections 
$C_{i,k}^{(S),(1)}$ there are again only single powers of $\ln (1-x)$, and 
Eq.~(\ref{ln1}) provides their correct analytic continuation. 

Combining everything, we arrive at the final result for $z<1$ for the NLO 
($\overline{\rm{MS}}$ scheme) non-singlet and singlet time-like splitting 
functions $P_{ij}^{(T),(1)}(z)$ and the NLO time-like short-distance cross 
sections $C_{i,k}^{(T),(1)}(z)$. The result for the NLO splitting functions
is in complete agreement with that of \cite{cfp,fp}, apart from a known 
misprint\footnote{The term $(10-18x-16 x^2/3) \ln x$ in the $C_F T_f$ part of 
$P_{qq}^{(T),(1)}(x)$ in \cite{fp} must correctly read \cite{grvfrag} 
$(-10-18x-16 x^2/3) \ln x$.} in \cite{fp}. The endpoint contributions, i.e., 
the terms $\sim \delta (1-z)$, to the diagonal splitting functions can be 
obtained from the fermion number and energy-momentum-conservation conditions 
\cite{cfp,fp} and are exactly the same as in the space-like situation.
In case of the NLO time-like quark short-distance cross section the endpoint 
contributions differ from those in the space-like situation by $C_F \pi^2 
\delta (1-z)$ which is just the effect of the above mentioned
$\pi^2$-correction when crossing the threshold at $Q^2=0$ \cite{cfp}. 
Taking this into account, the results for the $C_{i,k}^{(T),(1)}(z)$ agree
with those in, e.g., \cite{fpzp}\footnote{There is a typographical error
in the first equation of appendix II in \cite{fpzp}: the prefactor of the
$\ln x$ term should correctly read $3(1+x^2)/(1-x)$.}. 
Since all the unpolarized expressions have 
appeared in the literature we do not repeat them here.
We only note that, in contrast to the leading order (cf.\ Eq.~(\ref{GLR})), 
the NLO differences $P_{ij}^{(T),(1)}(\xi)-P_{ij}^{(S),(1)}(\xi)$ and 
$C_{q,k}^{(T),(1)}(\xi)-C_{q,k}^{(S),(1)}(\xi)$ are non-zero (note that here 
$\xi \leq 1$ in both the space-like and the time-like functions), i.e., 
in addition to the ACR the GLR is also broken beyond LO \cite{cfp}, as we 
anticipated in the introduction.
%
\section{The breaking of the ACR revisited}
%
Before addressing the polarized case which we are mainly interested in,
let us return for a moment to 
Eqs.~(\ref{ACRNLO2})-(\ref{unpdiff}). The 
rather simple structure of the $P_{ij}^{\epsilon,(1)}(z)$ and its 
transparent origin suggest that there could be a more straightforward 
way of linking the time-like and the analytically continued space-like NLO 
splitting functions, than going through Fig.~1 graph by graph and picking up
the higher pole terms. The starting point for such considerations is to
notice that Eq.~(\ref{ACRgen}) (when adapted to the {\em un}polarized case, 
i.e., with the $\Delta$'s omitted and the $P_{ji}^{(S),(0)}$ as given in
(\ref{ulosplit})) only states that the {\em four}-dimensional 
LO splitting functions obey the ACR. The rule must break down for the 
$(d=4-2 \epsilon)$-dimensional counterparts, 
$P_{ij,4-2\epsilon}^{(U),(0)}$, of 
the $P_{ij}^{(U),(0)}$ as an immediate consequence of Eq.~(\ref{ACRNLO}). We 
write down a LO analogue of Eq.~(\ref{ACRNLO2}) in $d$ dimensions,
\begin{equation} \label{ACRLO}
P_{ij,4-2 \epsilon}^{(T),(0)} (z) =  z {\cal AC} \Bigg[ 
P_{ji,4-2 \epsilon}^{(S),(0)} (x=\frac{1}{z}) \Bigg] + 
P_{ij}^{\epsilon,(0)}(z)  
\end{equation}
(for $z<1$), the main difference being that the LO $P_{ij}^{\epsilon,(0)}
(z)$ are only ${\cal O} (\epsilon)$ and not ${\cal O}(1)$:
\begin{eqnarray} \label{ACRLO1}
P_{qq}^{\epsilon,(0)}(z)&=& (-2 \ln z) \epsilon P_{qq}^{(S),(0)}(z) \; ,
\nonumber  \\
P_{gq}^{\epsilon,(0)}(z)&=& (-2 \ln z-1) \epsilon P_{gq}^{(S),(0)}(z) \; ,
\nonumber \\
P_{qg}^{\epsilon,(0)}(z)&=& (-2 \ln z+1) \epsilon P_{qg}^{(S),(0)}(z) \; ,
\nonumber \\
P_{gg}^{\epsilon,(0)}(z)&=& (-2 \ln z) \epsilon P_{gg}^{(S),(0)}(z) \; . 
\end{eqnarray}
As already seen from the example of Eq.~(\ref{icon}), the pieces $\sim 
\epsilon$ in the $d$-dimensional LO splitting functions result in finite 
contributions in the calculation of the {\em NLO} splitting functions.
One therefore suspects that the breakdown of the ACR beyond leading order in 
the $\overline{\rm{MS}}$ scheme, as expressed by 
Eqs.~(\ref{ACRNLO2})-(\ref{unpdiff}), is entirely driven by the corresponding 
breaking in the part $\sim \epsilon$ of the 
$d$-dimensional {\em LO} splitting 
functions, cf. Eqs.~(\ref{ACRLO}),(\ref{ACRLO1}). If this is indeed the case, 
then the functions $P_{ij}^{(T),(1)}$, $C_{i,k}^{(T),(1)}$ and 
\begin{eqnarray} \label{ptilde}
\tilde{P}_{ij}^{(T),(1)} &\equiv& z {\cal AC} \Bigg[ P_{ji}^{(S),(1)} 
(x=\frac{1}{z}) \Bigg] \; , \nonumber \\         
\tilde{C}_{i,k}^{(T),(1)} &\equiv& z {\cal AC} \Bigg[ C_{i,k}^{(S),(1)} 
(x=\frac{1}{z}) \Bigg] \; ,         
\end{eqnarray}
respectively, should be simply related by a factorization scheme 
transformation\footnote{For the non-singlet case this possibility was already
hinted at in \cite{cfp}.}. 
The general form of such a transformation can be determined from the condition
that is must leave any physical quantity such as, e.g., ${\cal{F}}_1^{(T)}$ or
${\cal{F}}_2^{(T)}$ invariant, and reads
\begin{eqnarray} \label{trafo}
P_{qq,\pm}^{(T),(1)} &\longrightarrow& P_{qq,\pm}^{(T),(1)} 
-\frac{\beta_0}{2} z_{qq}^{(T)} \; , \nonumber \\
\hat{P}^{(T),(1)} &\longrightarrow& \hat{P}^{(T),(1)}
-\frac{\beta_0}{2} \hat{Z}^{(T)} + \Bigg[ \hat{Z}^{(T)}, \hat{P}^{(T),(0)} 
\Bigg]_{\otimes}\; , \nonumber \\
C_{i,k}^{(T),(1)} &\longrightarrow& C_{i,k}^{(T),(1)} - z_{iq}^{(T)} \; ,
\end{eqnarray}
where the subscript '$\otimes$' denotes convolution when evaluating the 
commutator. Again, $\hat{P}^{(T),(0)}$ and $\hat{P}^{(T),(1)}$ are the 
(unpolarized) LO and NLO evolution matrices, respectively (cf. 
Eq.~(\ref{singmat})), and $z_{qq}^{(T)}$ and the $2\times 2$ matrix 
$\hat{Z}^{(T)}$ generate the transformation. In analogy with 
Eq.~(\ref{singmat}) we set\footnote{For our purposes, we do not need to 
distinguish between the non-singlet $z_{qq}^{(T)}$ and the $qq$ entry in 
the singlet matrix $\hat{Z}^{(T)}$ even though these could be chosen 
differently in principle.}
\begin{eqnarray} \label{zmat}
\hat{Z}^{(T)} \equiv \left( \begin{array}{cc} 
z_{qq}^{(T)} & 2n_f z_{gq}^{(T)} \\
\frac{1}{2n_f} z_{qg}^{(T)} & z_{gg}^{(T)} \\
\end{array}\right) \; . 
\end{eqnarray}  
According to Eq.~(\ref{ACRLO1}) one now expects that the choice 
\begin{equation} \label{zmat1}
z_{ij}^{(T)}(z) = (2 \ln z+a_{ij}) P_{ij}^{(S),(0)} (z) 
\end{equation}
with the logarithms being of kinematical origin and the $a_{ij}$ resulting 
from the adjustment of the spin-averaging factors, 
\begin{equation} 
a_{qq}=a_{gg}=0 \; , \;\;\; a_{gq}=1 \; , \;\;\; a_{qg}=-1 \; , 
\end{equation}
transforms all time-like NLO ($\overline{\rm{MS}}$) quantities to a scheme 
in which they satisfy the ACR, i.e., in which 
\begin{eqnarray} \label{ptilde1}
P_{ij}^{(T),(1)} &=& \tilde{P}_{ij}^{(T),(1)} = z {\cal AC} \Bigg[ 
P_{ji}^{(S),(1)} (x=\frac{1}{z}) \Bigg]  \; , \nonumber \\         
C_{i,k}^{(T),(1)} &=& \tilde{C}_{i,k}^{(T),(1)} = z {\cal AC} \Bigg[ 
C_{i,k}^{(S),(1)} (x=\frac{1}{z}) \Bigg] \; .         
\end{eqnarray}
This indeed turns out to be the case as one finds upon insertion of the 
$z_{ij}^{(T)}$ in (\ref{zmat1}) into Eq.~(\ref{trafo}) and comparison with 
(\ref{unpdiff}).
We emphasize that the space-like NLO quantities on the right-hand-sides
of Eq.~(\ref{ptilde1}) have not been transformed and are still in the 
$\overline{\rm{MS}}$ scheme. Eq.~(\ref{ptilde1}) therefore links quantities 
referring to different factorization schemes. This is perfectly legitimate
since one is free to choose the factorization schemes independently for
the space-like and time-like cases\footnote{For instance, one can choose to
factorize initial- and final-state collinear singularities differently in any 
higher order calculation.}. On the other hand, it does not really appear 
sensible from a physical point of view, and it actually turns out 
\cite{grvfrag} that the transformed time-like NLO splitting functions of
(\ref{ptilde1}) do no longer obey the energy-momentum-conservation condition. 
Anyway the above scheme transformation is not meant to be used in any 
practical calculation, it just serves to identify the breakdown of the ACR  
beyond LO as a mere matter of convention and provides a very transparent and 
remarkably simple way of obtaining the correct $\overline{\rm{MS}}$ time-like 
splitting functions from the analytically continued space-like ones. We note 
that in \cite{flor} the unpolarized NLO time-like splitting functions and 
short-distance cross sections were calculated using the cut vertex method.
In this formalism the validity of the ACR occurs quite naturally if certain 
renormalization conditions are imposed \cite{beau}, and the results of 
\cite{flor} therefore correspond to the $\tilde{P}_{ij}^{(T),(1)}$, 
$\tilde{C}_{i,k}^{(T),(1)}$ in (\ref{ptilde1}) rather than to the 
$\overline{\rm{MS}}$ scheme results.
%
\section{NLO results for the polarized case}
%
The extension of our results in sections 3 and 4 to the spin-dependent case 
is rather straightforward now. We first write down 
Eqs.~(\ref{ACRNLO2}),(\ref{ACRNLO2c}) for the polarized case,
\begin{eqnarray} \label{ACRNLO2p}
\Delta P_{ij}^{(T),(1)} (z) &=&  z {\cal AC} \Bigg[ \Delta P_{ji}^{(S),(1)}   
(x=\frac{1}{z}) \Bigg] + \Delta P_{ij}^{\epsilon,(1)}(z) \; , \\
\Delta C_{i}^{(T),(1)} (z) &=& z {\cal AC} \Bigg[ \Delta C_{i}^{(S),(1)}     
(x=\frac{1}{z}) \Bigg] + \Delta C_{i}^{\epsilon,(1)} (z) \; , 
\label{ACRNLO2cp}
\end{eqnarray}
where $z<1$ and where we have recalled from section 2 that contrary to the 
unpolarized case there is only one longitudinally polarized structure 
function for pure photon exchange, ${\cal G}_1^{(U)}$. For the space-like 
situation, NLO ($\overline{\rm{MS}}$) results for the short-distance cross 
sections $\Delta C_{i=q,g}^{(S),(1)}$ (i.e., the coefficient functions for 
$g_1^{(S)}$) have first been published quite some time ago \cite{kod,rat,bq}, 
whereas the corresponding $\overline{\rm{MS}}$ splitting functions have been 
calculated only fairly recently via the OPE \cite{vn} (where they appear as 
the anomalous dimensions) and in \cite{wv1,wv}, where the method of 
\cite{egmpr,cfp} was used. To be more precise, use of dimensional 
regularization in such NLO calculations for the polarized case implies
to choose a prescription for dealing with the Dirac matrix $\gamma_5$ and the 
Levi-Civita tensor $\epsilon_{\mu\nu\rho\sigma}$ in $d\neq 4$ dimensions,
which poses a non-trivial problem. In \cite{vn} the 'reading point'
prescription of \cite{korn} with a fully anticommuting $\gamma_5$ was chosen, 
whereas \cite{wv} adopted the original definition for $\gamma_5$ of 
\cite{hvbm} (HVBM scheme) which is widely considered to be the most consistent
method. A crucial feature in both \cite{vn} and \cite{wv1,wv} was that the 
genuine ('naive') $\overline{\rm{MS}}$ scheme result for the non-singlet NLO 
splitting function $\Delta P_{qq,+}^{(S),(1)}(x)$ (cf. Eq.~(\ref{ns2evol})) 
possessed the disagreeable property of having a non-zero first moment 
($x$-integral), in obvious conflict with the conservation of the non-singlet 
axial current \cite{kod1,svw} which demands that the first moment 
of the non-singlet quark combination $\Delta q_+^{(S)}$ be independent of 
$Q^2$. This effect was clearly due to the $\gamma_5$ prescriptions chosen and 
could be removed by a finite renormalization in \cite{vn} or, equivalently, 
by a factorization scheme transformation in \cite{wv1,wv} generated by 
the difference of the $d$-dimensional LO quark-to-quark splitting functions 
for the unpolarized and polarized (HVBM scheme) cases\footnote{As was also 
shown in \cite{wv1,wv,svw}, the scheme transformation corresponding to 
(\ref{pqqdim}) is needed at the same time to bring the first moment of the 
quark non-singlet coefficient function into agreement with the value given by 
the Bj\o rken sum-rule \cite{bjsr}.},
\begin{equation} \label{pqqdim}
\Delta P_{qq,4-2 \epsilon}^{(S),(0)} (x)-P_{qq,4-2 \epsilon}^{(S),(0)} (x) =
4 C_F \epsilon (1-x) \; .
\end{equation}
It turned out that both \cite{vn} and \cite{wv1,wv} then arrived at the same 
final result for the space-like polarized NLO splitting functions and also
for the coefficient functions $\Delta C_{q}^{(S),(1)}$ and $\Delta 
C_{g}^{(S),(1)}$ for which the previous results of \cite{kod,rat} and 
\cite{bq}, respectively, 
were confirmed. In a strictly technical sense of the word, the 
results of \cite{vn,wv1,wv} thus 
do not correspond to the $\overline{\rm{MS}}$ 
scheme. On the other hand, the '$\gamma_5$-effect' described above has 
been known to occur in the HVBM scheme for some time 
\cite{alex,arg2,svw,kamal}
and is purely artificial in the sense that it is related to helicity 
non-conservation at the quark-gluon 
tree-level vertex in $d\neq 4$ dimensions  
as expressed by the non-vanishing of the rhs of Eq.~(\ref{pqqdim}). Since 
physical requirements like the conservation of the non-singlet axial current 
serve to remove the effect in a straightforward and obvious way, the final 
results of \cite{vn,wv1,wv} are nevertheless usually regarded as the 
'real' conventional $\overline{\rm{MS}}$ scheme results. 

The reason for going into this discussion is the following. If we use the 
final results of \cite{vn,wv1,wv} for the $\Delta P_{ji}^{(S),(1)}$ and the 
$\Delta C_{i}^{(S),(1)}$ to obtain their time-like counterparts via
Eqs.~(\ref{ACRNLO2p}),(\ref{ACRNLO2cp}), the factorization scheme
transformation generated by (\ref{pqqdim}) and performed in the space-like
situation will obviously also affect the time-like result. On the other 
hand, in the case of the time-like NLO quantities, there appears to be no 
obvious requirement that enforces a certain value for, say, the first 
moment $\int_0^1 \Delta P_{qq,+}^{(T),(1)}(z) dz$. Thus in principle 
one would be allowed equally well to use the space-like 'naive' 
$\overline{\rm{MS}}$ scheme results in (\ref{ACRNLO2p}),(\ref{ACRNLO2cp}), 
i.e., those that possess the wrong (non-vanishing) value for the integral 
of $\Delta P_{qq,+}^{(S),(1)}(x)$. However, taking into consideration the 
origin of the above '$\gamma_5$-effect' as a pure artefact of the dimensional
calculation in a certain $\gamma_5$ prescription, we decide against this 
latter option and will use the final results of \cite{vn,wv1,wv}, i.e.,
the 'real' $\overline{\rm{MS}}$ scheme results for the space-like case
in what follows. This choice of factorization scheme appears most sensible
since it actually turns out that Eq.~(\ref{pqqdim}) remains completely 
unchanged when transformed to the time-like situation,
\begin{equation} \label{pqqtdim}
\Delta P_{qq,4-2 \epsilon}^{(T),(0)} (z)-P_{qq,4-2 \epsilon}^{(T),(0)} (z) =
4 C_F \epsilon (1-z) \; ,
\end{equation}
implying that the unphysical helicity non-conservation at the quark-gluon 
vertex in $d\neq 4$ dimensions also takes place in the time-like case if one 
uses the HVBM prescription for 
$\gamma_5$ \cite{arg}. Our choice obviously has 
implications for factorizing collinear singularities in NLO calculations of 
other cross sections with polarized final state particles: The 'real' 
$\overline{\rm{MS}}$ scheme factorization counterterm for all collinear poles 
coming from polarized (time-like) quark-to-quark transitions should be 
taken as \cite{arg}
\begin{displaymath}
-\frac{1}{\epsilon} \frac{\alpha_s}{2\pi} \left( \frac{M_F^2}{4\pi\mu^2} 
\right)^{-\epsilon} \frac{\Gamma (1-\epsilon)}{\Gamma (1-2 \epsilon)} 
\left[ \Delta P_{qq}^{(T),(0)}(z) + 4 C_F \epsilon (1-z) \right] 
\otimes \Delta \sigma^{LO}_{4-2 \epsilon} \; , 
\end{displaymath}
where $M_F$ is the factorization scale and $\Delta \sigma^{LO}_{4-2 \epsilon}$
is some appropriate polarized Born-level cross section in $d$ dimensions.

After these precautions we can turn to the analytic continuation to $x>1$ of 
the space-like NLO quantities which is required for 
Eqs.~(\ref{ACRNLO2p}),(\ref{ACRNLO2cp}) and works in exactly the same way as 
for the unpolarized case studied in section 3. The other ingredients to 
Eqs.~(\ref{ACRNLO2p}),(\ref{ACRNLO2cp}), $\Delta P_{ij}^{\epsilon,(1)}(z)$ 
and $\Delta C_{i}^{\epsilon,(1)} (z)$, are also straightforwardly calculated 
following the lines of section 3. The situation is facilitated by the fact 
that in the polarized case there are obviously no complications due to the 
gluon spin-averaging factors. Thus we only have to keep track of the 
$z^{-2 \epsilon}$ terms. We find:
\begin{eqnarray} \label{poldiff}
\Delta P_{qq,\pm}^{\epsilon,(1)}(z) &=& \beta_0 \Delta P_{qq}^{(S),(0)}(z)
\ln z = P_{qq,\pm}^{\epsilon,(1)}(z)\;\;, \nonumber \\
\Delta P_{qq,PS}^{\epsilon,(1)}(z) &=& - 12 C_F T_f \left[(1+z)\ln z + 
2(1-z)\right] \;\; , \nonumber \\
\Delta P_{gq}^{\epsilon,(1)}(z) &=& 4 (C_F-C_A) 
\Delta P_{gq}^{(S),(0)}(z)     
\left[-2 S_1(z)+\ln(1-z)\ln z \right] \nonumber \\
&+& C_F^2 \Bigg[ (4-5z)\ln z -4(1-z) \Bigg] 
+ 8 C_A C_F \Bigg[ (1+z)\ln z + 2 (1-z) \Bigg] \;\; , \nonumber \\
\Delta P_{qg}^{\epsilon,(1)}(z) &=& - \frac{8}{3} T_f \Delta P_{qg}^{(S),(0)}
(z) \ln z + 4 (C_F-C_A) \Delta P_{qg}^{(S),(0)} (z) \left[2 S_1(z)-
\ln(1-z)\ln z \right] \nonumber \\
&+& 2 C_F T_f \Bigg[ (5-4z) \ln z + 4(1-z) \Bigg] 
+ \frac{4}{3} C_A T_f 
\Bigg[ (10z-23) \ln z -24(1-z) \Bigg] \;\; , \nonumber \\
\Delta P_{gg}^{\epsilon,(1)}(z) &=& - \Delta P_{qq,PS}^{\epsilon,(1)}(z)
+ \beta_0 \Delta P_{gg}^{(S),(0)}(z) \ln z \;\; , \nonumber \\ 
\Delta C_{q}^{\epsilon,(1)} (z) &=& 2 \ln z \Delta P_{qq}^{(S),(0)}(z) \; , 
\nonumber \\
\Delta C_{g}^{\epsilon,(1)} (z) &=& 2 \ln z \Delta P_{gq}^{(S),(0)}(z) \; .
\end{eqnarray}

As for the unpolarized case in section 4 it turns out that all these
terms can also be fully accounted for by a factorization scheme 
transformation, i.e., the $\overline{\rm{MS}}$ 
scheme $\Delta P_{ij}^{(T),(1)}$
and $\Delta C_{i}^{(T),(1)}$ and the corresponding analytically continued
space-like functions, 
\begin{eqnarray} \label{dptilde}
\Delta \tilde{P}_{ij}^{(T),(1)} &\equiv& z {\cal AC} \Bigg[ \Delta 
P_{ji}^{(S),(1)} (x=\frac{1}{z}) \Bigg] \; , \nonumber \\         
\Delta \tilde{C}_{i}^{(T),(1)} &\equiv& z {\cal AC} \Bigg[ \Delta 
C_{i}^{(S),(1)} (x=\frac{1}{z}) \Bigg]\;,          
\end{eqnarray}
respectively, are related via Eqs.~(\ref{trafo}),(\ref{zmat}) 
(with $\Delta$'s everywhere in (\ref{trafo}),(\ref{zmat})) if one chooses
\begin{equation} \label{dzmat1}
\Delta z_{ij}^{(T)}(z) = 2 \ln z \Delta P_{ij}^{(S),(0)} (z) \; .
\end{equation}

Now we finally insert everything into Eqs.~(\ref{ACRNLO2p}),(\ref{ACRNLO2cp})
and arrive at the final results for the 
time-like NLO quantities which, in case
of the splitting functions, are conveniently expressed as differences with
respect to the space-like situation, 
at the same time indicating the breakdown 
of the GLR:
\begin{equation} \label{deldef}
\Delta P_{ij}^{(T),(1)} (z)=\Delta P_{ij}^{(S),(1)} (z) + \Delta_{ij}(z) \; ,
\end{equation}   
where the $\Delta P_{ij}^{(S),(1)} (z)$ are found in \cite{vn,wv1,wv} and 
\begin{eqnarray}
\label{polfin}   
\Delta_{qq,\pm}(z) &=& C_F^2 \frac{\ln z}{1-z} \left[ 4 \left(1+z^2\right) 
\ln (1-z) -
\left(1+3 z^2\right) \ln z + z^2+4z+1 \right] \; , \\
\Delta_{qq,PS}(z) &=& 4\,C_F T_f  \left( \ln z -3\right) \left[
(1+z)\ln z + 2(1-z)\right] \; , \\
\nonumber
\Delta_{gq}(z) &=& \frac{1}{3}\, C_F^2 \left[ (2-z) \left[-24 S_1(z)-4 \pi^2+
6 \ln^2(1-z)+12 \ln (1-z) \ln z \right. \right. \\
\nonumber
&& \left.\left. - 3 \ln^2 z + 15 \ln z \right] 
+ (15z-6) \ln(1-z)-33z+54 \right]\\
\nonumber
&+& \frac{1}{9}\, C_A C_F \left[6 (2-z) \left[12 S_1(z)+2\pi^2-3 \ln^2(1-z)
\right] \right. \\
\nonumber
&& \left. +(6-39 z) \ln(1-z) -18 (z+4) \ln^2 z +72 (3z-1) \ln z - 71 z 
+4 \right] \\
&+& \frac{4}{9}\,C_F T_f \left[ 3 (2-z) \ln(1-z) + z + 4 \right] \; , \\
\nonumber
\Delta_{qg}(z) &=& \frac{8}{9}\,T_f^2 \left[-3 (2z-1) (\ln (1-z)+
\ln z)-4z -1 \right] \\
\nonumber
&+& \frac{2}{3}\, C_F T_f \left[ (2z-1) 
\left[24 S_1(z)+4 \pi^2 - 6 \ln^2 (1-z) -3 \ln^2 z\right] \right. \\
\nonumber
&& \left. +(6z-15) \ln (1-z) + 30 \ln z - 78 z +57 \right] \\
\nonumber
&+& \frac{2}{9}\, C_A T_f \left[ 6(2z-1) \left[-12 S_1(z)-2 \pi^2+
6 \ln(1-z)\ln z \right.\right. \\
\nonumber
&& \left.\left. +3 \ln^2 (1-z)\right]-
(6z-39) \ln(1-z) +36 (1+z) \ln^2 z \right. \\
&& \left. - 3 (26z+11) \ln z + 284 z -217 \right] \; , \\
\nonumber
\Delta_{gg}(z) &=& 4\,C_F T_f \left[ 6(1-z) + 6 \ln z + (1+z) \ln^2 z
\right]\\
\nonumber
&-& \frac{8}{3}\,C_A T_f  \frac{2z^2-3z+2}{1-z} \ln z \\
\nonumber
&+& \frac{2}{3}\, C_A^2 \frac{\ln z}{1-z} \left[ 12 (2z^2-3z+2) 
\ln(1-z) + 6 (3z-4) \ln z \right.  \\
&& \left. -26z^2+63 z -26 \right] \; . 
\label{delgg}
\end{eqnarray}
For the short-distance cross sections we obtain
\begin{eqnarray}
\label{cqres}                 
\Delta C_q^{(T),(1)} (z) &=& C_F \left[ (1+z^2) \Bigg( \frac{\ln (1-z)}{1-z}
\Bigg)_+ + 2 \frac{1+z^2}{1-z} \ln z -\frac{3}{2} \frac{1}{(1-z)_+} 
+\frac{1}{2} (1-z) \right. \nonumber \\
&& \left. + \left( -\frac{9}{2} + \frac{2}{3} \pi^2 \right) \delta (1-z) 
\right]\; ,\\ 
\Delta C_g^{(T),(1)} (z) &=& C_F \left[ (2-z) \ln (z^2 (1-z)) -4 +3 z \right]
\; , \label{polfinc}   
\end{eqnarray}
in agreement with the results of \cite{arg} for the corresponding choices
$\Delta \tilde{f}_q^D (z) = -4 (1-z)$ and  $\Delta \tilde{f}_g^D (z) = 0$
in Eq.\ (14) of that paper\footnote{Note that our 
definition for the gluonic
short-distance cross section differs by a factor of 2 from the one
used in \cite{arg}.}. The '+'-prescription in (\ref{cqres}) 
is defined as usual via
\begin{equation}
\int_0^1 dz f(z) \left( g(z) \right)_+ \equiv 
\int_0^1 dz \left(f(z)-f(1)\right) g(z)\:\:.
\end{equation}

For numerical evaluations of our results it is convenient to have the 
Mellin-moments of the expressions above which are defined by 
\begin{equation} \label{mellin}
f[n] \equiv \int_0^1 z^{n-1} f(z) dz
\end{equation}   
and are presented in the appendix. Fig.~3 provides a comparison of our results
for the NLO ($\overline{\rm{MS}}$ scheme) time-like polarized and unpolarized
splitting functions in Mellin-$n$ space. 
\begin{figure}[hbt]
\begin{center}
\vspace*{-1.5cm}
\epsfig{file=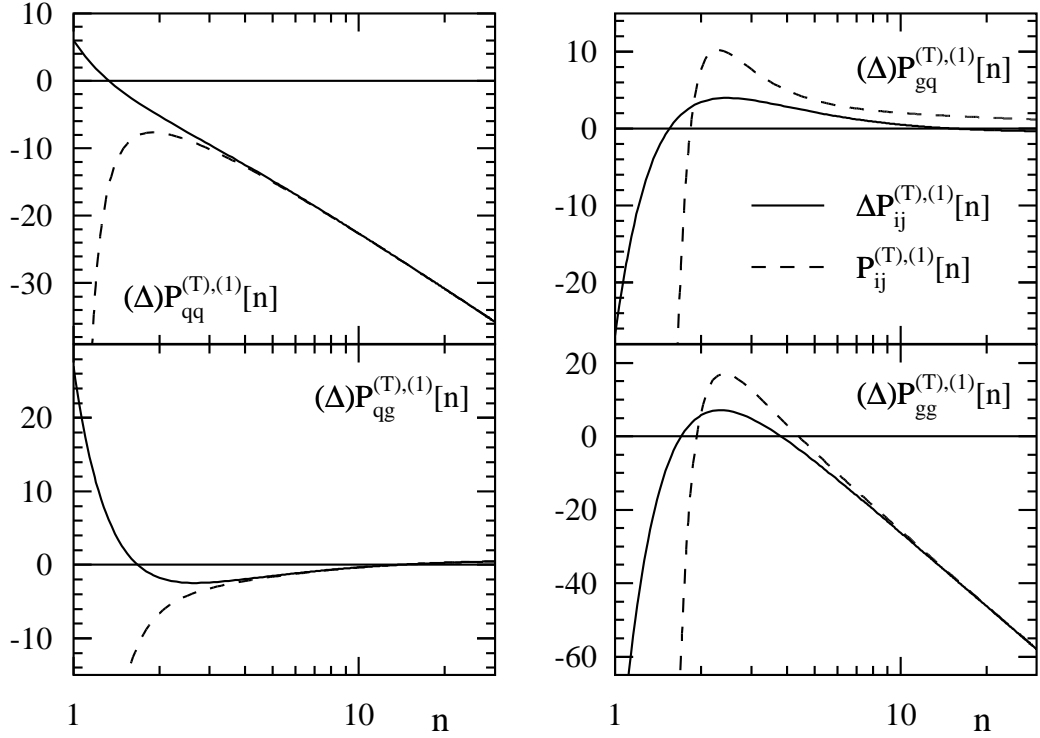,height=12cm}
\vspace{-1.3cm}
\caption{Comparison of the spin-dependent NLO ($\overline{\rm{MS}}$) 
time-like singlet splitting functions $\Delta P_{ij}^{(T),(1)}[n]$ as 
functions of Mellin-$n$ according to Eqs.\ (A.1)-(A.6) with the corresponding
unpolarized ones, as taken from [\ref{grvfr}], for $f=3$ flavors.}
\end{center}
\vspace{-0.5cm}
\end{figure}
It is interesting to observe that all LO and NLO 
time-like splitting functions
obey $\Delta P_{ij}^{(T),(k)}[n] \rightarrow P_{ij}^{(T),(k)}[n]$ 
($k=0,\,1$; $i,\,j=q,\,g$) 
as $n\rightarrow \infty$, i.e., as $z\rightarrow 1$, except
for $\Delta P_{gq}^{(T),(1)}[n]$.
A similar observation was made for the space-like quantities where, again,
only $\Delta P_{gq}^{(S),(1)}[n]$ does not fulfil
$\Delta P_{ij}^{(S),(k)}[n] \rightarrow P_{ij}^{(S),(k)}[n]$
as $n\rightarrow \infty$ \cite{grsv}.
Finally, the values for the first ($n=1$) 
moments of the polarized NLO singlet quantities are given by
\begin{eqnarray}
&&\Delta P_{qq}^{(T),(1)} [1] = 3 C_F T_f  \;\; , \;\;\;  
\Delta P_{gq}^{(T),(1)} [1]= -3 C_A C_F +C_F^2 \left( \frac{3}{2} -\pi^2 
\right) \; , \nonumber \\ 
&&\Delta P_{qg}^{(T),(1)} [1] = 2 T_f \beta_0  \;\; , \;\;\; 
\Delta P_{gg}^{(T),(1)} [1] = -\frac{5}{6} C_A^2 -4 C_F T_f -\frac{1}{3} 
C_A T_f - \frac{1}{3} \pi^2 C_A \beta_0 \; , \nonumber \\ 
&&\Delta C_q^{(T),(1)} [1] = 0  \;\; , \;\;\; 
\Delta C_g^{(T),(1)} [1] = -\frac{29}{4} C_F \; .
\end{eqnarray}
%
\section{A supersymmetric property of the NLO time-like splitting functions}
%
In this section we finally very briefly address a 
relation that is conjectured 
for an ${\cal N}=1$ supersymmetric Yang-Mills theory and connects all singlet 
splitting functions in a remarkably simple way in the limit $C_F=N_C=2 T_f
\equiv N$ (cf.\ \cite{dok}). In, e.g., the unpolarized case it reads          
\begin{equation} \label{susy}
\delta [\hat{P}^{(U),(i)}(\xi)] \equiv P_{qq}^{(U),(i)}(\xi) + 
P_{gq}^{(U),(i)} (\xi) - P_{qg}^{(U),(i)}(\xi) - P_{gg}^{(U),(i)}(\xi) \equiv 
0 \; .
\end{equation}
In LO ($i=0$), the relation is satisfied by the both the unpolarized and the
polarized splitting functions and, trivially, in both the space-like and
the time-like situations. Beyond LO, one can expect it to continue to hold
{\em only} if the regularization method adopted respects supersymmetry.
One therefore anticipates that the NLO ($\overline{\rm{MS}}$) space-like
and time-like splitting functions of dimensional regularization will
not satisfy (\ref{susy}), in agreement with the findings in \cite{fp,vn}.
However, one regularization method that is applicable to supersymmetry is 
dimensional {\em reduction} \cite{siegel}, a variant of dimensional 
regularization. The scheme essentially consists of performing the 
Dirac-algebra in {\em four} dimensions and of continuing only momenta to 
$d$ ($d<4$) dimensions. In order to match the ultraviolet (UV) sectors of 
dimensional regularization and dimensional reduction, specific counterterms 
need to be introduced \cite{ssk,kt} in the latter which include a finite 
renormalization of the strong charge. Once this is done, all remaining 
differences between the results for a NLO quantity in dimensional 
regularization and in dimensional reduction can only be due to the effects of 
mass singularities. They are fully accounted for \cite{kt,kst,kamal}
by the differences between the $d$-dimensional LO splitting functions  
(as to be obtained in dimensional {\em regularization}) and the 
{\em four}-dimensional ones (corresponding to dimensional {\em reduction}).
In other words, the breakdown of the supersymmetric relation for dimensional
regularization is entirely blamed on the breakdown of this relation in the 
$\sim \epsilon$-parts of the $d$-dimensional LO splitting functions of 
dimensional regularization. This feature was exploited in \cite{kst,wv} to
transform the space-like unpolarized and polarized NLO splitting functions of 
$\overline{\rm{MS}}$ dimensional 
regularization to dimensional reduction via a 
simple factorization scheme transformation and to establish the validity of 
Eq.~(\ref{susy}) for the obtained quantities\footnote{The supersymmetry 
relation for the space-like NLO kernels was proved prior to \cite{kst,wv}
in the OPE calculations of \cite{af,vn} in which the transition from 
dimensional regularization to dimensional reduction occurs as a finite 
renormalization.}. We will now extend the considerations of \cite{kst,wv}
to the time-like situation. In the unpolarized case, one finds for the NLO 
$\overline{\rm{MS}}$ splitting functions of dimensional regularization
in the limit $C_F=N_C=2 T_f \equiv N$ \cite{fp}:
\begin{equation} \label{susyunp}
\delta [\hat{P}^{(T),(1)}(z)] = N^2 \left[ -\frac{2}{3z}+\frac{13}{6} 
+\frac{5}{3} z-z^2 +(-1+2 z+4 z^2) 
\ln z -\frac{1}{2} \delta (1-z) \right] 
\end{equation}                                        
where we have included the endpoint contribution.
The corresponding expression for the polarized NLO splitting functions 
can be obtained from Eqs.~(\ref{deldef})-(\ref{delgg}) 
and the space-like results of
\cite{vn,wv1,wv},
\begin{equation} \label{susypol}
\delta [\Delta \hat{P}^{(T),(1)}(z)] = N^2 \left[ \frac{11}{6}-\frac{7}{6}z 
+(1-z) \ln z -\frac{1}{2} \delta (1-z) \right] \; .
\end{equation}                                        
What we need to do is to perform factorization scheme transformations 
(\ref{trafo}) of these results, with the $(\Delta ) z_{ij}^{(T)}$ to be 
determined from the parts $\sim \epsilon$ of the time-like $d$-dimensional LO 
splitting functions as obtained in dimensional regularization. For our 
purposes we only need to consider (\ref{trafo}) in the combination appearing 
on the lhs of Eq.~(\ref{susy}) and in the supersymmetric limit:
\begin{equation} \label{trafosusy}
\delta [(\Delta ) \hat{P}^{(T),(1)}] \longrightarrow \delta [ (\Delta )
\hat{P}^{(T),(1)}] - \frac{\beta_0}{2} \delta [(\Delta ) \hat{Z}^{(T)}] +
\left( (\Delta )P_{qg}^{(T),(0)} + (\Delta ) P_{gq}^{(T),(0)} \right) \otimes 
\delta [ (\Delta ) \hat{Z}^{(T)}] \; , 
\end{equation}
where now $\beta_0=3 N$. The calculation of the parts $\sim \epsilon$ in the 
$d$-dimensional LO time-like splitting functions yields\footnote{As
discussed in the previous section the
difference (\ref{pqqtdim}) between the LO polarized and unpolarized time-like 
quark-to-quark splitting functions arising in the HVBM scheme is already 
accounted for in Eqs.~(\ref{polfin})-(\ref{polfinc}). To obtain the result in 
(\ref{zsusypol}) one therefore has to use $\Delta P_{qq,4-2 \epsilon}^{(T),
(0)} (z)=P_{qq,4-2 \epsilon}^{(T),(0)} (z)$ (see also \cite{wv}).}
\begin{eqnarray} \label{zsusy}
\delta [ \hat{Z}^{(T)} (z)] &=& N \left( 1-2 z +2 z^2 -\frac{1}{3} 
\delta (1-z) \right) \; , \\
\delta [ \Delta \hat{Z}^{(T)} (z)] &=& N \left( 1-z - \frac{1}{3} 
\delta (1-z) \right) \; . \label{zsusypol}
\end{eqnarray}
Upon insertion of Eqs.~(\ref{susyunp}),(\ref{zsusy})  (or 
(\ref{susypol}),(\ref{zsusypol})) into (\ref{trafosusy}) one finds that the 
resulting transformed $\delta [\hat{P}^{(T),(1)}]$ (or $\delta [\Delta 
\hat{P}^{(T),(1)}]$, respectively) vanishes identically. We thus have 
demonstrated the validity of the 
${\cal N}=1$ supersymmetric relation also for 
the time-like NLO polarized and unpolarized evolution kernels in dimensional 
reduction. Apart from being interesting of its own, this finding also
provides evidence for the correctness of our results in sections 3-5.
%
\section{Summary}
%
We have presented a calculation of the unpolarized and polarized time-like 
NLO splitting functions, needed for the NLO $Q^2$-evolution of 
(spin-dependent) fragmentation functions. The starting point for our 
considerations were \cite{cfp,fp,ev,wv1,wv} in which the 
corresponding space-like quantities were calculated within a method based on 
the factorization properties of mass singularities in the light-cone gauge. 
As was shown in \cite{cfp} for the non-singlet case one can then determine 
the time-like counterparts via analytic continuation to $x>1$, which is 
also the way we have pursued. It turned out that beyond LO there are certain 
terms arising from phase space (and, for the unpolarized case, from the gluon 
spin-averaging in $d\neq 4$ dimensions) which prevent the analytic 
continuation relation (ACR) of \cite{dly} between the space-like 
and time-like splitting 
functions from remaining intact. The same statement applies to the connection 
between the space-like and time-like short distance cross sections of
electroproduction and $e^+ e^-$ annihilation, respectively. Nevertheless, 
the corrections to the ACR are rather straightforwardly calculable within the 
method of \cite{cfp}. Even more, we were able to show that 
in both the unpolarized and the polarized cases one can transform the results
to a factorization scheme, different from the 
$\overline{\rm{MS}}$ scheme, in which the breakdown of the ACR does not
occur. In the unpolarized case our final 
$\overline{\rm{MS}}$ results confirm those of 
\cite{cfp,fp} obtained within the same method, whereas in the polarized case
our results are entirely new. Finally we have shown that, when transformed
to dimensional reduction, both our unpolarized and polarized results for the 
time-like NLO splitting functions satisfy a simple relation motivated from 
supersymmetry.
\section*{Acknowledgements}
The work of M.S.\ has been 
supported in part by the 'Bundesministerium f\"{u}r 
Bildung, Wissenschaft, Forschung und Technologie', Bonn.
\section*{Appendix}
\setcounter{equation}{0}
\renewcommand{\theequation}{\rm{A}.\arabic{equation}}
\vspace*{-0.6cm}
In this appendix we present the Mellin-$n$ moments (as defined in 
Eq.~(\ref{mellin})) of our NLO results for the polarized case in 
Eqs.~(\ref{polfin})-(\ref{polfinc}). 
The corresponding $n$ moments for the unpolarized time-like splitting
functions and short-distance cross sections can be found in \cite{grvfrag}.
As in (\ref{deldef}) we write 
\begin{equation} 
\Delta P_{ij}^{(T),(1)} [n]=\Delta P_{ij}^{(S),(1)} [n] + \Delta_{ij}[n] \; ,
\end{equation}   
where the $\Delta P_{ij}^{(S),(1)} [n]$ can be found in a compact analytic
form\footnote{Note that the results presented in \cite{grsv} need to be 
divided by $-8$ in order to bring them to our normalization for the NLO 
splitting functions. Also note that the definition of the $\pm$-components
in the non-singlet sector (see Eqs.~(\ref{nsdef})-(\ref{ns2evol})) occurs in 
a reversed notation in \cite{grsv}, i.e., as $\mp$.} in \cite{grsv} and the 
$\Delta_{ij}[n]$ are given by:
\begin{eqnarray}
\Delta_{qq,\pm}[n] &=& C_F^2 \left[4 {\cal S}_1(n)-\frac{3n^2+3n+2}{n(n+1)}
\right] \left[ -2 {\cal S}_2(n)+ \frac{\pi^2}{3}+\frac{2n+1}{n^2(n+1)^2}
\right] \; , \\
\Delta_{qq,PS}[n] &=& 4\,C_F T_f  \frac{(n+2)(3n+1)}{n^3 (n+1)^3} \; , \\
\nonumber
\Delta_{gq}[n] &=& C_F^2 \left[ -2 \frac{n+2}{n(n+1)} 
\left(-{\cal S}_1^2(n)+{\cal S}_2(n)+\frac{\pi^2}{3}\right) - 
\frac{3n^3+n^2-18n-8}{n^2(n+1)^2}{\cal S}_1(n) \right. \\
\nonumber
&& \left. + \frac{7n^5+22n^4+7n^3-24n^2-22n-4}{n^3(n+1)^3}\right]\\
\nonumber
&+& C_A C_F \left[-2 \frac{n+2}{n(n+1)}({\cal S}_1^2(n)-3 {\cal S}_2(n)) +
\frac{(n+2)(11n-1)}{3n(n+1)^2} {\cal S}_1(n) \right. \\
\nonumber
&& \left. - \frac{\left(67n^4+101 n^3 + 34n^2 +144n+72\right)
\left(n+2\right)}{9n^3 (n+1)^3}
\right]\\
&+& \frac{4}{9}\, C_F T_f \left[-3 \frac{n+2}{n(n+1)} {\cal S}_1(n) +
\frac{(n+2)(5n+2)}{n(n+1)^2} \right] \; , \\
\nonumber
\Delta_{qg}[n] &=& \frac{8}{9}\,T_f^2 \left[ 3 \frac{n-1}{n(n+1)} 
{\cal S}_1(n) - \frac {5n^3-3n^2+7n+3}{n^2 (n+1)^2}\right]\\
\nonumber
&+& 2\,C_F T_f \left[2 \frac{n-1}{n(n+1)} (-{\cal S}_1^2(n)+3 {\cal S}_2(n)) +
\frac{3n^2+5}{n(n+1)^2} {\cal S}_1(n)\right. \\
\nonumber
&& \left. -\frac{7n^5+7n^4-5n^3+5n^2+4n-2}{n^3 (n+1)^3}\right]\\
\nonumber
&+& \frac{2}{9}\, C_A T_f \left[6\frac{n-1}{n(n+1)}(3 {\cal S}_1^2(n)-3
{\cal S}_2(n)-\pi^2) \right. \\
\nonumber
&& \left. - 3 \frac{11n^3-12n^2+37n+12}{n^2(n+1)^2} {\cal S}_1(n) \right. \\
&& \left. + \frac{67n^5+34n^4+32n^3+98n^2+249n+72}{n^3(n+1)^3}\right] \; , \\
\Delta_{gg} [n] &=& -8 \,C_F T_f \frac{n^3+3n^2-1}{n^3(n+1)^3} 
+ \frac{4}{3}\,C_A T_f \left[ -2 {\cal S}_2(n) + \frac{\pi^2}{3} + 
4 \frac{2n+1}{n^2(n+1)^2} \right] \nonumber \\
&+& C_A^2 \left[4 {\cal S}_1(n)-\frac{11n^2+11n+24}{3 n (n+1)} \right]\;
\left[-2 {\cal S}_2(n) +\frac{\pi^2}{3} + 4 \frac{2n+1}{n^2(n+1)^2}\right] 
\end{eqnarray}
with 
\begin{equation}
{\cal S}_k (n) \equiv \sum_{j=1}^n \frac{1}{j^k} \; .
\end{equation}
The analytic continuation of the 
${\cal S}_k$, required for a numerical Mellin 
inversion, is well-known \cite{grvmom}. For the moments of the polarized NLO
time-like short-distance cross sections we find
\begin{eqnarray} 
\Delta C_q^{(T),(1)}[n] &=& C_F \Bigg[ {\cal S}_1^2 (n) + 5 {\cal S}_2 (n)
+\left( \frac{3}{2} -\frac{1}{n (n+1)} \right) {\cal S}_1 (n) +
\frac{3}{(n+1)^2} -\frac{1}{2 (n+1)} \nonumber \\
&& -\frac{1}{n}-\frac{2}{n^2} -\frac{9}{2} \Bigg] \; , \\
\Delta C_g^{(T),(1)}[n] &=& C_F \Bigg[ -\frac{2+n}{n (n+1)} {\cal S}_1 (n) 
-\frac{4}{n} -\frac{4}{n^2} +\frac{3}{n+1} +\frac{3}{(n+1)^2}  \Bigg] \; .
\end{eqnarray}
%

%
\end{document}